\documentclass[aps,prl,groupedaddress,twocolumn,showpacs]{revtex4-1}
\usepackage{graphicx}
\usepackage{latexsym}
\usepackage{amsmath}
\usepackage{txfonts}
\usepackage{helvet}
\usepackage{braket}

\begin{document}

\title{Symmetry Preservation and Critical Fluctuations in a Pseudospin Crossover Perovskite LaCoO$_3$}

\author{Yasuhiro Shimizu$^{1,*}$, Takuya Takahashi$^1$, Syunpei Yamada$^1$, Ayako Shimokata$^1$, Takaaki Jin-no$^2$, Masayuki Itoh$^1$}

\affiliation{$^1$Department of Physics, Nagoya University, Furo-cho, Chikusa-ku, Nagoya 464-8602, Japan}
\affiliation{$^2$Technical Center, Nagoya University, Furo-cho, Chikusa-ku, Nagoya 464-8601, Japan}
\email[email:]{yasuhiro@iar.nagoya-u.ac.jp}
\date{\today}

\begin{abstract}
Spin-state crossover beyond a conventional ligand-field theory has been a fundamental issue in condensed matter physics. Here, we report microscopic observations of spin states and low-energy dynamics through orbital-resolved NMR spectroscopy in the prototype compound LaCoO$_3$. The $^{59}$Co NMR spectrum shows the preserved crystal symmetry across the crossover, inconsistent with $d$ orbital ordering due to the Jahn-Teller distortion. The orbital degeneracy results in a pseudospin ($\tilde{J} = 1$) excited state with an orbital moment observed as $^{59}$Co hyperfine coupling tensors. We found that the population of the excited state evolves above the heart crossover temperature. The crossover involves critical spin-state fluctuations emerging under the magnetic field. These results suggest that the spin-state crossover can be mapped into a statistical problem, analogous to the supercritical liquid in liquid-gas transition. 
\end{abstract}

\maketitle



	Paramagnetic local moments in condensed matter systems typically lose their entropy via magnetic ordering and spin-singlet formations by breaking the time-reversal and translational symmetries at low temperatures. An exceptional case is the spin-state crossover (SC) induced by the ligand field that overcomes Hund's exchange coupling in transition metal ions \cite{Goodenough}. The realization of unusual spin states due to many-body correlations remains controversial \cite{Korotin}. In a prototype compound LaCoO$_3$, the low-spin (LS, $t_{2g}^6$, $S = 0$) ground state of Co$^{3+}$ under the octahedral ligand field $\Delta_{\rm oct} = 10Dq$ changes into a paramagnetic state favored by Hund's coupling with increasing temperature $T$ (Fig. 1) \cite{Goodenough}. The paramagnetic moment (e. g., that evaluated from the multiplication of magnetic susceptibility $\chi$ and $T$) continuously increases above 30 K and is further enhanced above $500$ K where the system becomes itinerant [Fig. 1(b)]. Although the paramagnetic state was initially considered as a high-spin state (HS; $S$ = 2, $e_g^2t_{2g}^4$) \cite{Goodenough}, the proposal of an intermediate-spin state (IS; $S$ = 1, $e_g^1t_{2g}^5$) stabilized by $d$ orbital ordering \cite{Korotin} has triggered extensive research [Fig. 1(c,d)] \cite{Maris, Asai, Yamaguchi, Zobel, Ishikawa, Klie, Phelan}. However, further experiments including the neutron diffraction \cite{Radaelli} and inelastic scattering measurements \cite{Podlesnyak} provide a negative evidence for IS. This contradiction implies the limitation of a single-ion picture neglecting the thermal population change \cite{Miyashita} and the spin-orbital entanglement. As reflected by a degenerate ground state, the LS state transitions into a magnetic state under high magnetic field \cite{Sato, Altarawneh, Ikeda}, hole doping\cite{Itoh2000}, and strain \cite{Fuchs}. 
	
\begin{figure}
	\includegraphics[scale=0.88]{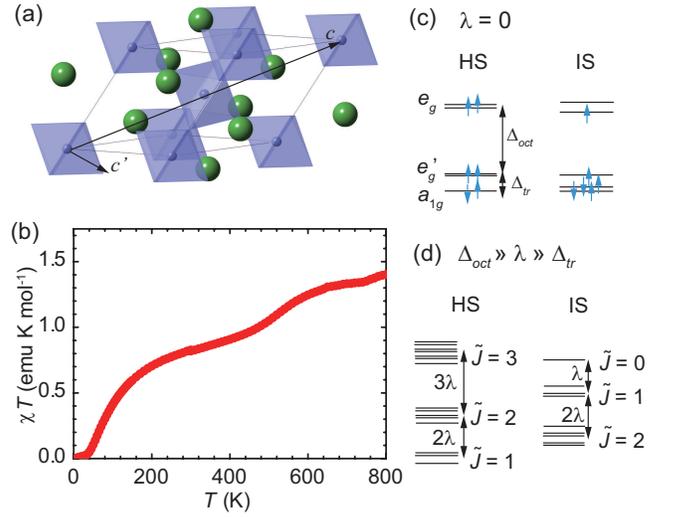}
	\caption{\label{Fig1} 
	(a) Crystal structure of LaCoO$_3$ (space group: rhombohedral $R{\bar{3}}c$). Long and short arrows denote the $c$ axis of a hexagonal lattice and the cubic axis tilted from the $c$ axis by 54.7$^\circ$, respectively. (b) Magnetic susceptibility $\chi$ multiplied by the temperature $T$ as a function of $T$. (c) $3d$ orbital levels are mainly split into a $e_g$ doublet and a $t_2g$ triplet by $\Delta_{\rm oct} \sim 0.5$ eV. The latter is lifted into $e_g^\prime$ and $a_{1g}$ under the trigonal ligand field $\Delta_{\rm tr}$, where the occupations for HS and IS refers to the density functional theory calculation \cite{Hsu}. (d) $\tilde{J}$-multiplet energy schemes in the intermediate ligand field regime, $\Delta_{\rm oct} \gg \lambda \gg \Delta_{\rm tr}$, with $\tilde{L} = 1$. HS and IS have $(2S + 1)(2\tilde{L} + 1)$ = 15 and 9 manifolds, respectively, which are split by $\lambda $ into $\tilde{J} = 1, 2, 3$ and $\tilde{J} = 0, 1, 2,$ respectively \cite{Ropka, Haverkort, Podlesnyak, SI}. 
	}
\end{figure}
	One of the key experimental strategies for discriminating between the HS and IS states is to observe symmetry lowering due to the Jahn-Teller distortion expected in IS \cite{Korotin}. Another is determining the $t_{2g}$ orbital occupation that differs between the HS and IS states in band structure calculations \cite{Hsu, Eder, Harrison}. For the $t_{2g}$ triplet with a small trigonal field $\Delta_{\rm tr} \sim 7$ K, the spin-orbit coupling ($\lambda = -145$ cm$^{-1}$ = $-208$ K) well exceeds $\Delta_{\rm tr}$ (but is smaller than $\Delta_{\rm oct}$) \cite{Itoh, Noguchi, Ropka, Haverkort, Podlesnyak}. In the intermediate ligand-field regime \cite{Abragam}, the good quantum number should be replaced by the effective angular momentum or pseudospin $\tilde{J}$ with a pseudo-orbital moment $\tilde{L} = 1$ of the $t_{2g}$ triplet. The observed $g$ value ($g_\parallel = 3.35$ and $g_\perp = 3.55$) \cite{Noguchi} and the magnetic circular dichroism \cite{Haverkort} indicate a significant spin-orbit coupling. In that case, the lowest multiplet level is $\tilde{J} = 1$ for HS [Figs. 1(c) and 1(d)] \cite{Abragam, Haverkort}. The pseudo spin state of IS can be $\tilde{J} = 0$ or $\tilde{J} = 2$ for $\tilde{L} = 1$ \cite{Haverkort, SI}. 

	To uncover the local spin state and low-lying excitations across SC, we have conducted NMR measurements on a single crystal of LaCoO$_3$. Improving upon the previous work on powder \cite{Itoh} and twinned crystal samples \cite{Kobayashi}, our extensive single-crystal experiments elucidate the thermal variation of excited-state populations via the anisotropic hyperfine interactions probed by the Knight shift $K$ and the nuclear quadrupole frequency $\nu_{\rm Q}$ measurements. The dynamical properties of the electronic and the lattice degrees of freedom are independently investigated by $^{59}$Co and $^{139}$La nuclear spin relaxation rates. 

Single crystals of LaCoO$_3$ were grown by a floating zone method (3 mm/hr) under a O$_2$ pressure of 7 atm. The crystal axes for cleaved crystal faces were determined from the Laue diffraction patterns. The frequency-swept $^{59}$Co and $^{139}$La NMR spectra were obtained from spin-echo signals after $\pi/2$-$\tau$-$\pi/2$-$\tau$ pulses ($\pi/2$ = 1.5 $\mu$s, $\tau = 4-20$ $\mu$s) with 300 kHz steps at a constant magnetic field of 6.10 T. The nuclear spin-lattice relaxation time, $T_1$, was measured for the magic angle of 54.7$^\circ$ measured from the $c$ axis, where the nuclear magnetization recovery follows a single exponential function. 


\begin{figure}
	\includegraphics[scale=0.65]{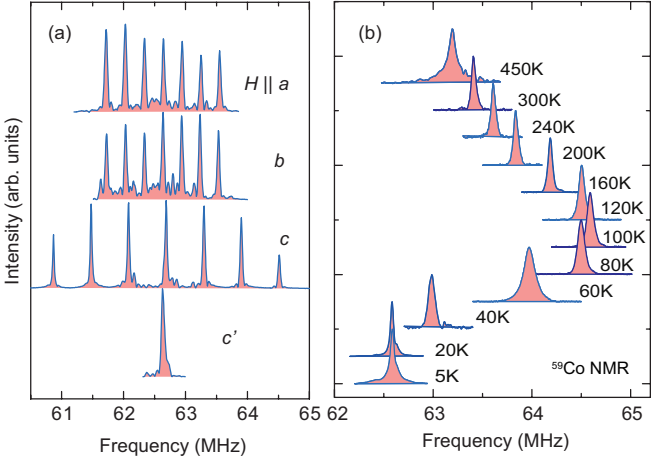}
	\caption{\label{Fig2} 
	(a) The $^{59}$Co NMR spectra of the low-spin state at 20 K for magnetic field directions along crystal axes ($a$, $b$ and $c$) and the magic angle ($c^\prime$). (b) Temperature dependence of the $^{59}$Co NMR spectrum for $H \parallel c^\prime$, where the intensity is normalized to the peak value at each temperature. 
	}
\end{figure}
	First, we examine the crystal symmetry of LaCoO$_3$ from the angular dependence of the $^{59}$Co NMR spectrum. For a rhombohedral $R\bar{3}c$ lattice \cite{Radaelli}, all Co sites are equivalent in arbitrary field directions. Indeed, we find that the $^{59}$Co spectrum simply consists of seven equally spaced lines due to quadrupole splitting of the $^{59}$Co nuclear spin $^{59}I = 7/2$, as shown in Fig. 2(a). The split frequency $\delta \nu$ has a maximum (defined as $^{59}\nu_{\rm Q}$) at the $c$ axis and is independent of the field direction ($^{59}\nu_{\rm Q}/2$) in the $ab$ plane. The axial anisotropy reflects the symmetry of the electric field gradient (EFG) due to the trigonal distortion of CoO$_6$ along the $c$ axis [Fig. 1(a)]. $\delta \nu$ vanishes at a magic angle $c^\prime = 54.7^\circ$, which is the so-called cubic axis, measured from the $c$ axis. 

By increasing $T$ from 5 to 450 K, the $^{59}$Co NMR spectrum ($H \parallel c^\prime$) shows a paramagnetic Knight shift with a maximum around 100 K [Fig. 2(b)]. The spectral width remains sharp ($< 0.2$ MHz) without quadrupole splitting and site doubling, which are expected for a monoclinic lattice distortion due to orbital ordering \cite{Maris}. This provides the evidence for the preserved crystal symmetry of $R\bar{3}c$ in the extensive temperature range of $5$--$450$ K. In addition, it rules out the possibility of static heterogeneity such as a LS and HS order \cite{Kunes, Werner, Suzuki}. 
	
	\begin{figure*}
	\includegraphics[scale=0.78]{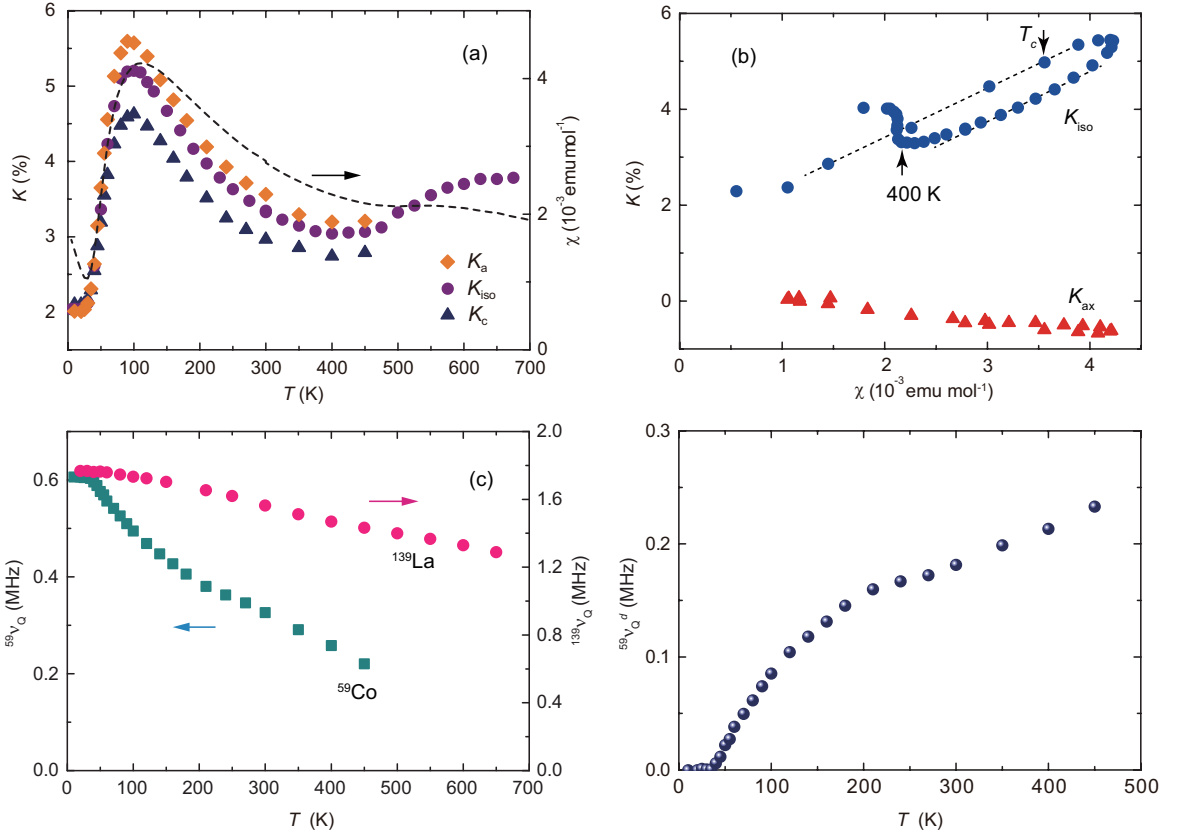}
	\caption{\label{Fig3} 
	(a) Temperature dependence of $^{59}$Co Knight shift $K$ for the $c$ and $c^\prime$ axes (the left-hand axis), in comparison with the magnetic susceptibility $\chi$ (dashed curve, the right-hand axis). $K_{\rm iso}$ was obtained for a powdered sample above 450 K. (b) $K - \chi$ plots for the isotropic and axial parts of $^{59}$Co Knight shifts. (c) $^{59}$Co (the left-hand axis) and $^{139}$La (the right-hand axis) quadrupole frequency $^i \nu_Q$ ($i$ = 59 and 139). (d) Temperature dependence of the nuclear quadrupole frequency $^{59}\nu_Q^d$ that measures the HS population. 
	}
	\end{figure*}
Figure 3(a) shows the temperature dependence of the Knight shifts ($K_{c^\prime}$, $K_a$, $K_c$) for $H \parallel c^\prime$, $a$, and $c$ axes, respectively, where $K_{c^\prime}$ is equivalent to the isotropic shift $K_{\rm iso} = (2K_a + K_c)/3$ for the axial symmetry. Unlike typical paramagnets, the $K_{\rm iso}$--$\chi$ plot exhibits a strong nonlinear behavior [Fig. 3(b)]. A hyperfine constant $A_{\rm iso}$ of 5.8 T/$\mu_{\rm B}$ is obtained for limited temperature ranges 30--60 K and 160--260 K. As the hyperfine interaction consists of the orbital term $\mathcal {P} {\bf \tilde{L}} \cdot {\bf I}$ ($ > 0$), the spin term $-\mathcal {P}\kappa {\bf S} \cdot {\bf I}$ ($ > 0$), and a small Van Vleck term ($\mathcal{P}$ a nuclear dependent constant, $\kappa$ a core polarization coefficient) \cite{SI}, positive $A_{\rm iso}$ represents the predominant orbital term. 

Across the SC that involves level crossing of the HS and the LS states, the Van Vleck susceptibility $\chi_{\rm VV} = N\mu_{\rm B}^2\Lambda_{pq}$ strongly depends on the temperature, where $\Lambda_{pq} = \sum_n\frac{\bra{O} L_p \ket{n}\bra{n} L_q \ket{O}}{E_n-E_0}$ for energy levels $E_n$ and $E_0$ corresponding to the excited states $\ket{n}$ and the nondegenerate ground state $\ket{O}$ respectively. As $\chi_{\rm VV}$ is related to the Van Vleck Knight shift $K_{\rm VV}$ via the different hyperfine coupling constant 2$\mathcal{P}$, the nonlinear $K$--$\chi$ relation appears around $T_c = 50$ K [Fig. 3(b)], where $E_{\rm LS}$ becomes lower than $E_{\rm HS}$ (see also Fig. S2 \cite{SI}). $K_{\rm VV}$ is also enhanced above 400 K, where the temperature scale is comparable to the energy difference between $\tilde{J} = 1$ and $\tilde{J} = 2$ ($2\lambda \sim 416$ K), giving rise to the second SC. 

The anisotropic part of the hyperfine coupling constant, $A_{\rm ax}$, reflects a spin-polarized $3d$ orbital. We obtained $A_{\rm ax} = -1.6$ T/$\mu_{\rm B}$ from the axial Knight shift $K_{\rm ax} = 2(K_c - K_a)/3$ plotted against $\chi$ [Fig. 3(b)]. As the upper $e_g$ occupation remains degenerate under the trigonal field, $A_{\rm ax}$ is governed by the occupation difference between $e_g^\prime$ and $a_{1g}$ among the $t_{2g}$ triplet, which depends on the spin state [Fig. 1(c)] \cite{Hsu}. The negative $A_{\rm ax}$ is consistent with the dipole hyperfine interaction with a predominant $e_g^\prime$ spin \cite{Abragam, Shimizu}. It suggests that an excess electron out of the half-filling shell occupies into the $a_{1g}$ orbital. 

A more direct probe of the $d$ cloud distribution is the electric quadrupole interaction between the nuclear quadrupole moment and the EFG of the anisotropic $d$ orbitals with principal component, $V_Z^d$. The observed $^{59}\nu_Q$ consists of the outer-ion term, $^{59}\nu_Q^{\rm out}$, and the $d$ electron term, $^{59}\nu_Q^d$. $^{59}\nu_Q^{\rm out}$ can be evaluated through a point-charge approximation based on the crystal structure \cite{Radaelli}. Indeed, the $^{139}$La quadrupole frequency $^{139}\nu_Q$ governed by the outer-ion term [Fig. 3(c)] scales well to the calculated $^{139}\nu_Q$ (Fig. S3) \cite{SI}. We obtained $^{59}\nu_Q^d$ as $^{59}\nu_Q - (1 - \gamma_\infty)^{59}\nu_Q^{\rm out}$ with an appropriate antishielding factor $1 - \gamma_\infty = 7$. As shown in Fig. 3(d), $^{59}\nu_Q^d$ increases above 40 K, indicating the evolution of anisotropic $d$ orbital occupations. The smaller $^{59}\nu_Q(T)$ than $^{59}\nu_Q^{\rm out}$ signifies a sign difference between $V_Z^d$ and $V_Z^{\rm out}$. Here $V_Z^d$ should be negative, because the EFG of the outer ions $V_Z^{\rm out}$ is positive. The result is consistent with the excess $a_{1g}$ occupation, as expected in the HS state for the density functional calculation without spin-orbit coupling \cite{Hsu}. On the other hand, the IS state is expected to have an excess $e_g^\prime$ occupation giving positive $V_Z^d$. 
	

	\begin{figure*}
	\includegraphics[scale=0.85]{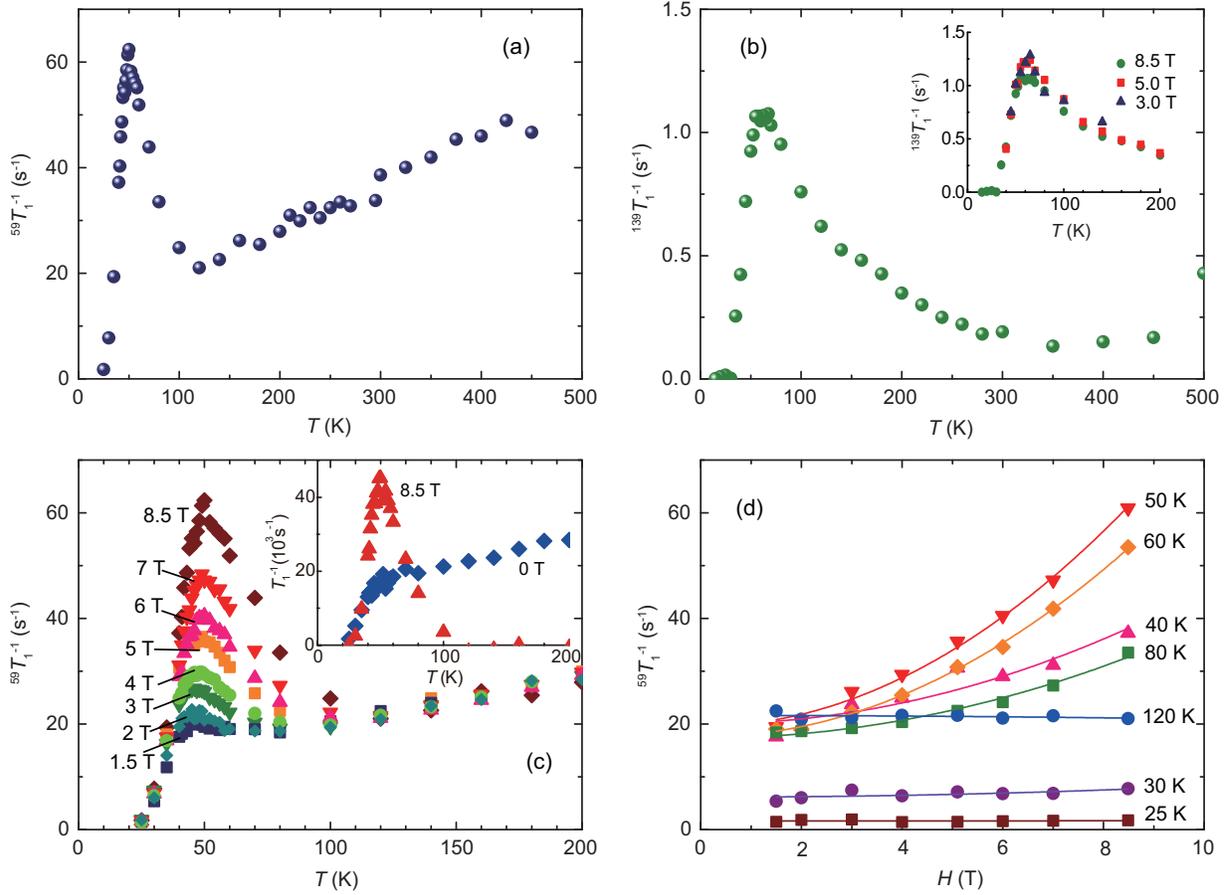}
	\caption{\label{Fig4}
Temperature dependence of (a) the nuclear spin-lattice relaxation rates $^{59}T_1^{-1}$ and (b) $^{139}T_1^{-1}$. (b) inset and (c) respectively show $^{139}T_1^{-1}$ and $^{59}T_1^{-1}$ at several magnetic fields below 200 K. (d) Magnetic field $H$ dependence of $^{59}T_1^{-1}$. Solid curves: fitting data with a parabolic function $^{59}T_1^{-1} = AH^2 + B$ for constants $A$ and $B$. }
	\end{figure*}
Low-energy excitations are investigated with the nuclear spin-lattice relaxation rates $^iT_1^{-1}$, $i$ = 59 and 139 for $^{59}$Co and $^{139}$La sites, respectively. As shown in Fig. 4(a), $^{59}T_1^{-1}$ exhibits a prominent peak around $T_c$, where $\chi$ has an inflection point ($d^2\chi / dT^2  = 0$, Fig. S4 \cite{SI}), and a minimum around 100 K. In contrast, $^{139}T_1^{-1}$ exhibits a considerably broader peak around $T_c$ and continues to decrease up to 300 K [Fig. 4(b)]. Such a strong site dependence of $^iT_1^{-1}$ is reminiscent of antiferromagnetic fluctuations, where $^iT_1^{-1}$ depends on the form factor $A(q)$ \cite{Takigawa}. In LaCoO$_3$, however, antiferromagnetic fluctuations have not been observed by inelastic neutron scattering measurements \cite{Asai, Podlesnyak}. Rather, the site dependence is attributed to the difference in the nature of the hyperfine interactions of the $^{59}$Co and $^{139}$La sites. More specifically, $^{139}T_1^{-1}$ for the closed shell La site with a larger electric quadrupole interaction is governed by the EFG fluctuations from the surrounding ions and is sensitive to the lattice softening. An anomaly around $T_c$ has also been observed in thermal expansion \cite{Zobel, Yan} and ultrasound measurements \cite{Naing}. 
	
	The origin of the $^{59}T_1^{-1}$ peak is further investigated by measuring the magnetic field $H$ dependence of $^iT_1^{-1}$, as shown in Fig. 4(c). We find a remarkable $H$ dependence in $^{59}T_1^{-1}$ around $T_c$. Upon lowering $H$, $^{59}T_1^{-1}$ is suppressed without changing the peak temperature and the peak almost vanishes at 1.5 T. In stark contrast, $^{139}T_1^{-1}$ displays no appreciable $H$ dependence [Fig. 4(b) inset]. In Fig. 4(d) we plot $^{59}T_1^{-1}$ as a function of $H$ at each temperature below 120 K. Between 30 and 80 K, it is well fitted to a parabolic function, $^{59}T_1^{-1} = AH^2 + B$, with fitting parameters $A$ and $B$. The $H$-independent component ($B$) evolves above 30 K [Fig. 4(c) inset]. The temperature dependence similar to $^{59}\nu_{\rm  Q}^d$ indicates that $B$ is governed by the HS population. As $K$ is independent of $H$ in the measured range (Fig. S5 \cite{SI}), the $H$-dependent component of $^{59}T_1^{-1}$ ($AH^2$) is not attributed to a field-induced population change which requires significantly higher fields ($> 30$ T) \cite{Sato, Altarawneh, Ikeda}. Therefore, the $^{59}T_1^{-1}$ peak at $T_c$ signifies a purely dynamical origin due to HS--LS fluctuations. 

	\begin{figure}
	\includegraphics[scale=0.7]{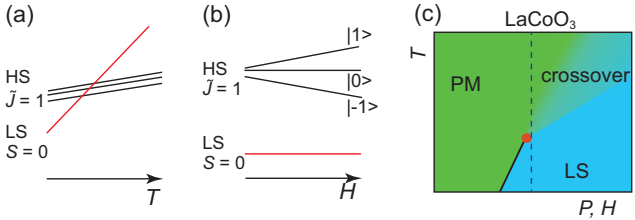}
	\caption{\label{Fig5} 
	(a) Level crossing scheme of pseudospin crossover between HS ($\tilde{J} = 1$) and LS ($S=0$). 
	(b) Zeeman splitting of the HS and LS energy levels.
	(c) Schematic phase diagram as a function of pressure $P$ or magnetic field $H$ versus temperature $T$, where the first-order transition line between paramagnetic (PM) and low-spin phases is terminated at a critical point. 
	}
	\end{figure}
To explain the field-dependent $^{59}T_1^{-1}$ peak, we consider a simple two-level system with a ground state $\ket{LS; S = 0}$ and an excited state $\ket{HS; \tilde{J} = 1}$, as shown in Fig. 5(a). $^{59}T_1^{-1}$ measures the electronic transition probability between the two levels as $\frac{2\pi}{\hbar }|\bra{LS}\mathcal{H^\prime}\ket{HS}|^2\delta (E_{LS} - E_{HS} + \hbar \omega_n)$, where the perturbed Hamiltonian $\mathcal{H^\prime} = -\gamma _n\hbar {\bf I \cdot \delta H}$ is the interaction with the hyperfine field $\delta {\bf H}$. $E_{LS}$ ($E_{HS}$) is the LS (HS) energy and $\omega_n$ is an NMR frequency. The LS and HS levels come close to each other at $T_{\rm c}$, i .e., $E_{LS} - E_{HS} \simeq 0$, as $E_{LS}$ decreases with the increasing ligand field upon cooling [Fig. 5(a)]. Thus, the low-energy transition probability is enhanced at $T_c$. However, a triplet degeneracy of $\ket{HS; \tilde{J} = 1}$ ($\tilde{J_z} = \pm1$ and 0) requires no net $\delta H$ observed by $^{59}T_1^{-1}$ in the absence of an external magnetic field. As the magnetic field lifts the triplet degeneracy through Zeeman splitting, $\delta H$ increases linearly with $H$ in the first order [Fig. 5(b)], resulting in a $H^2$ dependence of $^{59}T_1^{-1}$ around $T_{\rm c}$. Such low-energy fluctuations detected around the NMR frequency ($\sim$MHz) window differs from the energy scale of trigonal-field splitting within $\ket{HS; \tilde{J} = 1}$, as observed by EPR and inelastic neutron scattering experiments in the GHz range \cite{Noguchi, Podlesnyak}. 

	It is noted that $^{59}T_1^{-1}$ does not display a divergent behavior at $T_c$, unlike the critical fluctuations due to second-order magnetic transition. This highlights the ``crossover" character of the phase transition that preserves the lattice symmetry, reminiscent of liquid-gas transition. Thus, the above level crossing scheme of single ions oversimplifies the real system, requiring thermodynamic fluctuations and strong intersite correlations. In comparison with $^{59}T_1^{-1}$ and $^{139}T_1^{-1}$, the HS-LS fluctuations start to slow down only below 100 K, while lattice softening starts significantly earlier at temperatures below 300 K. The difference suggests that the SC is triggered by the dynamical lattice distortions. 

Theoretically, SC can be simply described by an Ising-like model with intersite elastic interactions between the LS and HS molecules of different sizes \cite{Miyashita, Konishi}. By systematically changing the spring constant, the first-order transition becomes a crossover with increasing pressure above the critical end point \cite{Konishi}. One can construct the pressure--temperature phase diagram of the SC system, analogous to the liquid--gas transition, as shown schematically in Fig. 5(c). Above the critical point, the spin state involves a dynamical admixture of the LS and HS states, which is regarded as a supercritical liquid of the liquid-gas transition. The observed sharp $^{59}T_1^{-1}$ peak implies that LaCoO$_3$ is located close to the critical point at ambient pressure. Indeed, SC becomes broader away from the critical point by applying pressure \cite{Asai2}. To investigate the criticality, which is expected to belong to a mean-field universality class \cite{Miyashita}, we have to apply negative pressure. Instead, the field-induced LS-HS transition may provide a possible way for accessing the critical point \cite{Altarawneh, Ikeda}. 
	 
	To summarize, we have shown that NMR distinguishes the spin state through the analysis of hyperfine interactions in LaCoO$_3$. Through anisotropic hyperfine interactions, we track the behavior of the $d$ electron occupation giving rise to the high-spin excited state. The absence of symmetry lowering from the $R\bar{3}c$ lattice rules out the possibility of $d$ orbital ordering due to the Jahn-Teller distortion. Based on the electric field gradient due to $d$ orbitals, thermal evolution of the HS population is elucidated across the spin crossover. The hyperfine interactions and the field-dependent fluctuations are a signature of the $\tilde{J} = 1$ pseudospin state. The contrasting magnetic and electronic fluctuations highlight the crucial mechanism of spin crossover driven by elastic interactions. Thus, our results provide deep microscopic insights into the historical spin crossover problem in prototype LaCoO$_3$.

	We thank S. Inoue for technical assistance, and S. Sugai and M. Murakami for experimental support in taking a Laue picture. This work was financially supported by KAKENHI Grants No. JP16K13836 and No. JP16H04012 from JSPS, and No. JP17H05151 from MEXT.

\end{document}